**Local political control in educational policy: Evidence from decentralized teacher pay reform under England's local education authorities**


Yiang Li[a,*], Xingzuo Zhou[b]

*Corresponding author

[a*] Division of the Social Sciences, University of Chicago, 1126 E 59[th] Street, Chicago, IL 60637, United States, +44 7529911038, yiangli@uchicago.edu
[b] Department of Economics, University College London



Funding: This research did not receive any specific funding from agencies in the public, commercial, or not-for-profit sectors.




**Abstract**

In 2012, the School Teachers' Review Body discontinued central guidance and allowed school discretion in determining teachers' pay in England. Meanwhile, local education authorities (LEAs) offer non-statutory teacher pay recommendations to LEA-controlled schools. This study examines how LEAs' political party control determines their guidance regarding whether to adopt flexible performance pay or continue seniority-based pay. A regression discontinuity design is used to address the endogeneity of political control and educational policy-making. We find that marginally Conservative-controlled LEAs are more inclined to recommend market-oriented flexible pay structures. The results remain robust to alternative specifications. This study reveals that politics matter in England's local educational policy-making, which has broad implications for future policy.

*Keywords*: Education Policy, Local Policy, Pay Decentralization, Public Management, UK Politics



**1. Introduction**

Good teachers are essential for high student attainment, and their impact continues into adulthood (Chetty et al., 2014; Rothstein, 2010). Research shows that decentralized teachers' pay and performance pay are critical for retaining high-quality teachers (Biasi, 2021; Britton & Propper, 2016; Burgess et al., 2019; Nasim et al., 2022). Moreover, the British public appreciates the significance of education as a public policy priority, and 42% believe that teacher pay is inadequate given their high workloads (Albakri et al., 2020). Before 2013, the School Teachers' Review Body (STRB), which advises the British government on teacher pay, followed an inflexible bureaucratic model in which teachers were paid based on their experience rather than performance (i.e., workloads and productivity). The seniority-based pay schedule accounts for the unmeasurable teacher output. However, it is insufficient to incentivize teachers' efforts and is unresponsive to the local labor market in competing for high-quality teachers (Department for Education, 2012; Rothstein, 2015).

During the 2013–14 academic year, STRB launched a reform of the teacher pay schedule and permitted schools controlled by local education authorities (LEAs) to adopt a performance-based flexible pay structure within the legally set maximum and minimum bounds. Under the reform, effective for teachers in 2014, schools referred to teachers' annual performance reviews to determine their pay. Furthermore, STRB stopped publishing the seniority-based pay spine points. As the reform offered schools a choice without mandating them to adopt flexible performance pay, the National Union of Teachers (union) continued to provide "shadow" seniority-based spine points for schools that followed the seniority-based pay schedule.

Meanwhile, LEAs publish non-statutory guidance documents every year advising LEA-controlled schools on teacher pay. Among the 99% of LEA-controlled schools that revised their pay policies, 84% relied on LEA pay recommendations (Burgess et al., 2019).



Although the documents contain pay recommendations for both classroom teachers and leadership staff, this study focuses on the former. LEA coordinates and balances the conflicting demands of multiple regional education stakeholders, including headteachers, unions, central government, parents (i.e., voters), and students (Eyles et al., 2017; Fitz, 2003). Thus, local policymakers face the difficult choice of either adopting flexible performance pay introduced by the central reform or following the seniority-based "shadow-STRB" union teacher pay recommendations.

In choosing teacher pay recommendations, elected LEA councilors rely on their party affiliation. Conservatives, compared to other British parties (e.g., Green and Labour), have a solid ideological reputation for adopting non-interventionist roles and fostering free-market mechanisms (Exley & Ball, 2011; Kenyon, 1995; Machin & Vignoles, 2006). Representing the capitalist class education ideology, Conservatives have a strong impulse toward school autonomy and using private-sector influence to achieve "academic rigor" (Jones, 2019). Their attempt to deregulate education is evident in 21st-century academization that has offered state schools autonomy outside of local government control (Hill et al., 2016). In 1990, the Conservative proposal to privatize state schools into academies was implemented by Tony Blair's Labour party for a small proportion of failing schools (Gill & Janmaat, 2019). However, the Conservatives' rapid expansion of academies to 75% secondary and 25% primary schools is not supported by today's Labour, Green, or Liberal Democrat parties (Crawford et al., 2020; Damm & Dayson, 2019). Hence, the significant shifts in partisan standpoints on market-based education policies, particularly in England, increase the relevance of understanding the effect of party control on teacher pay recommendations.

There is a lack of consensus in the research regarding the effects of political party control in a democratic majoritarian government on policy-making (Mair, 2008). The limited quantitative studies, predominantly in the US, have focused explicitly on education policies



under local partisan influence. Kitchens (2021) used multi-level, fixed-effect regression on aggregated precinct-level voting data to understand the effect of partisan support on US education funding within a school district. He found that controlling for state-level political control, a 1% increase in school district Democratic vote share caused a $9.59 increase in funding. In addition to *partisan support*, local *partisan control* in the US for both governors and the state legislature also influence educational policy-making. Chin and Shi (2021) studied the effect of the Democratic party's *partisan control* in the US state legislatures on education spending from 1984 to 2013, using a regression discontinuity design (RDD). Their results suggested that Democratic states spent 6.5% less than their Republican counterparts on K–12 education and, acting congruently with low-income supporters, traded off education for welfare spending on income redistribution. Using RDD, Ortega (2020) found similar results in US higher education, which witnessed 6% higher funding under Democratic governors than under their Republican counterparts. Few studies in non-US contexts have demonstrated partisan influence on education policies (e.g., Alves et al., 2021; Rauh et al., 2011).

In the UK, particularly England, the contexts of education and political institutions are different from in the US. The central UK government distributes local education funding only via LEAs with rules attached. Thus, LEAs have little power apart from keeping a small proportion to cover administrative costs (Agyemang, 2010). Existing research on the effect of political control in the UK on educational policy has primarily focused on how LEAs exercise flexibility in academization (i.e., privatizing state schools into independent academies). Gill and Janmaat (2019) used qualitative comparative analysis to perform mixed-methods research on the LEA factors associated with school academization. Their results suggested that high degrees of academization were associated with Conservative-controlled local authorities, especially those with underperforming schools in urban regions or those



disfavor private education. Academization empowers schools to flexibly set teacher pay, and it deviates from union recommendations. The aforementioned study by Gill and Janmaat (2019) suggested potential evidence of a connection between party affiliation and local government teacher pay guidance for LEA-controlled, non-academy schools. Other studies have widely noted that Conservatives show strong impulses in their policy-making toward market-oriented school autonomy and private-sector influence (e.g., Hill et al., 2015; Jones, 2019).

The aforementioned studies have made significant contributions toward understanding the effects of political control on policy-making. However, gaps remain in the existing evidence base. First, most studies have focused on public policies related to budgetary spending and social services contracting while neglecting social policies such as those related to education. Second, existing studies on the politics of educational policies, particularly in England, have rarely used quasi-experimental approaches (e.g., Agyemang, 2009; Grosvenor & Myers, 2006) that could causally attribute policy differences to political ideology. Third, the findings of most quasi-experimental studies conducted in the US and the rest of Europe have little generalizability to other political settings such as the UK. Fourth, while most existing studies use longitudinal data, they have not considered the political control of local governments over the years, which can also affect policy-making.

This study contributes to the literature on the partisan political drivers behind England's educational policy-making by using quasi-experimental methods related to the STRB educational reform. It uses LEA teacher pay recommendations from 2014 to 2017, collected via Freedom of Information (FOI) requests. Moreover, it links recoded recommendations to political party control over the years to shed light on the political motivations behind recommendations. Using RDD allows us to narrowly examine the differences between LEAs under the control of Conservatives or others, as well as whose



party affiliation is quasi-random. In Section 3.1, We discuss the detailed RDD model settings and the political institutions in England that allow for using the described methods.

This study aims to understand how LEA political control affects the flexibility of teacher pay recommendations to local schools. The research question is: Do Conservative-controlled LEAs have a higher probability of recommending that schools deviate from the union and adopt performance-based pay? The findings would contribute to the field of literature on the local political control behind policymaking and advise future effective central governance.

The rest of this paper is organized as follows. Section 2 describes the multiple data sources and provides the descriptive statistics. Section 3 outlines the empirical strategy. Section 4 presents the RDD results along with robustness checks while Section 5 discusses the findings and concludes the paper.



**2. Data**

*2.1 Data Sources and Overview*

We investigate the effect of LEA political party control on the flexibility of teacher pay recommendations. To this end, this study builds a repeated cross-sectional dataset that spans 2013–2017 and covers three key categories: teacher pay recommendations, election data, and other local LEA-level characteristics.

The first category of data describes the flexibility of teacher pay recommendations drawn from non-statutory "model school pay policies" issued by 152 LEAs to all LEA-maintained schools from 2013 to 2017. These are unclassified administrative documents collected via FOI requests sent in February 2020. Using the 16 objective questions given in Appendix A, two research assistants (RAs) and the author independently analyzed the collected LEA policy documents as part of a research project led by Nasim et al. (2022) at the UCL Social Research Institute, aiming to classify the flexibility of the LEA teacher pay recommendations. In the event of different judgments between the RAs and myself, we reconciled the responses and recoded them in a simplified spreadsheet. Finally, the responses to questions 1–8 were used to create a binary variable capturing the flexibility of LEAs' recommended pay policies regarding the union "shadow-STRB" recommendations. The final non-response rate was 19.4%. These were LEAs who did not respond to FOI requests despite multiple follow-ups or were unable to retrieve the documents because of administrative errors.

The second data category comprises the by-year political composition of English councils since 2000, including London's boroughs and unitary and county councils, compiled by Thrasher and Rallings (2022) from the Election Center at the University of Plymouth. The dataset was collected from the electoral commissions and local authority council administrative records. At the LEA level, these data delineate the type of governing council,



number of seats controlled by each party, total council seats, and the election outcomes. Throughout the analysis, the UK council election boundaries have remained the same under the Fifth Periodic Review in England, effective since 2007 (UK Parliament, 2007). In the dataset, there were missing election results for the City of London and the Isles of Scilly, both 100% controlled by independents without partisan affiliation throughout 2013–2017. We manually appended official election results collected from council websites to the dataset.

The final data section captures local council demographics and economic characteristics before the study period, drawn from the March 2011 UK Census (Office for National Statistics, 2011). The data were collected through 25 million pre-addressed questionnaires using a "specially developed national address register" across England and Wales (Office for National Statistics, 2013, p. 3). This also includes council-level crime administrative data recorded by the UK police in 2010 (Home Office, 2010) along with official England life expectancy data for 2013 (Office for National Statistics, 2020). When merging with the LEA data, mean imputation was used for missing income, crime rates, and life expectancy statistics (missing rates: 4.69%, 21%, and 4.69%, respectively).

*2.2 Sample Selection*

Initially, there were 760 LEA-year observations (i.e., 152 LEAs * 5 years). LEAs started advising schools on whether they should exercise flexibility in teacher pay in 2014. Thus, we excluded 2013, when none but the Conservative-controlled West Berkshire recommended flexible pay in anticipation of the upcoming reform (152 LEA-year observations dropped). From this, we further excluded 118 LEA-year observations (38 Conservative-controlled and 80 other-controlled LEAs) that did not answer FOI requests or responded that they could not retrieve the documents. There was no statistically significant political motivation behind the non-responses, controlling for year- and LEA-fixed effects (Conservative control dummy-parameter estimate: $\beta = 0.0012, p = 0.973$).



The final primary analytic sample includes 490 LEA-year observations of political composition and teacher pay recommendations (Table 1). In England, local government council elections are held on a fixed electoral cycle in May (Rallings & Thrasher, 2007). During the analysis period, from 2014 to 2017, every LEA held at least one election (UK Government, 2019).

| Sample selection criteria | Sample size | % Of full sample |
| --- | --- | --- |
| Full sample | 760 | 100% |
| Remove observations in 2013 when the reform did not officially start | 608 | 80% |
| Remove observations unable to supply pay document | 565 | 74% |
| Remove observations did not respond to FOI request | 490 | 64% |

**Table 1.** Sample selection table

*2.3 Variables*

The dependent variable is the flexibility of LEA teacher pay recommendations. The reconciled responses to questions 1–8 are used to answer the question "Was the local education authority advising schools differently from the union recommendation?" with a binary variable. The outcome of 1, following union recommendation, includes instances where LEAs recommended the same as or at most one spine point different from the union; LEAs explicitly asked schools to follow union recommendation and provided six spine points without specifying pound values. Then, from 2015, approximately 20% of LEAs every year recommended seven pay points, which, compared with the six recommended by the union, included an extra pay point of a 1% upgrade relative to the maximum in the previous year other than the union's 2% upgrade of the maximum. These seven-pay-point LEAs are not considered a separate category in our binary definition of the flexibility of teacher pay



recommendations and are merged with the outcome of 1. The outcome of 0, which deviates from the union, includes LEAs with six spine points including more than one difference from union recommendations, LEAs with more than six spine points excluding those with seven pay points, LEAs providing only maximum and minimum spine points or advising schools to follow STRB, and LEAs providing pay policy documents but having no recommendation inside. Alternative versions of the dependent variable are not chosen since the binary form allows the simplest modeling to reveal the causal relationship. Appendix D presents a multinomial logistic regression (MLR) on the three-category variable that disaggregates seven pay points from aligning with union recommendations as robustness checks.

Table 2 (panel A) shows the distribution of the flexibilities in LEA pay recommendations based on the controlling party in the primary full analytic sample. We examine the difference in the group means of the proportion of advising schools to follow union recommendations between Conservative-controlled and other-controlled LEAs. The naïve party difference suggests that the Conservative-controlled LEAs were less likely to advise schools to follow union recommendations in the primary full sample. However, this result is not statistically significant (Conservative control dummy-parameter estimates $\beta =$ 0.497, p = 0.203).



| | Full sample | | |
|---|---|---|---|

*Panel A. Flexibility of LEA teacher pay recommendation (binary) (%)*

| | Conservative LEAs (N=157) | Others-controlled LEAs (N=333) | Differences p-value |
|---|---|---|---|
| LEA teacher pay recommendation (follow union recommendation = 1, deviate from union recommendation = 0) | 0.497 | 0.559 | 0.203 |

*Panel B. Political Composition (%)*

| | 2014 | 2015 | 2016 | 2017 |
|---|---|---|---|---|
| Conservative LEAs | 35 | 36 | 40 | 47 |
| Conservative average council seats | 34.330 | 35.034 | 35.537 | 39.013 |
| Labour LEAs | 58 | 57 | 58 | 57 |
| Labour average council seats | 46.000 | 46.706 | 46.625 | 44.498 |
| Liberal Democrats LEAs | 1 | 1 | 1 | 1 |
| Liberal Democrats average council seats | 9.469 | 8.745 | 8.679 | 8.323 |
| No Overall Control (NOC) | 22 | 27 | 25 | 20 |
| Independent | 1 | 1 | 1 | 1 |

*Note:* The raw percentages do not include triangular kernel weighting, council size weighting, year fixed effect, and the LEA fixed effect. The elected councilors' party affiliation only corresponds to the time when the election took place, and, on rare occasions, there are elected politicians defecting their original party but mostly become independent; the descriptive statistics for the primary full-sample do not include the excluded LEA-year observations unresponsive to FOI or could not retrieve the pay document; hence the summary statistics do not fully reflect the England political landscape (e.g.., only 1 out of 2 independent LEAs are present because the others did not respond to FOI and are excluded from the primary full dataset).

**Table 2.** Variable distributions

The by-year party compositions of LEAs are used to create two variables. First, we derive the forcing variable, the margin of victory, defined by the percentages of LEA council



seats one party holds above or below the 50% majority threshold. This variable is used to derive the bandwidth that subsets a close-election sample for RDD analysis, which we discuss further in Section 3.1. The use of percentages ensures consistency of comparison across LEAs with different total council seats. Then, based on the victory margin, we also derive the treatment dummy variable of political party control by Conservatives or not. Furthermore, we rename the councils in the party composition dataset to ensure the names are consistent with the outcome measure before merging them. The renaming process is described in Appendix B.

Table 3 (panel B) shows the yearly political party compositions in the primary full sample. Over the years, the Conservatives gained control of more LEAs. Their average council seats also steadily increased relative to other political parties within the analysis period.

## 2.4 Descriptive Statistics

Table 4 provides the summary descriptive statistics of all 490 LEA-year observations by political party control in the repeated cross-sectional dataset. The baseline characteristics were collected from 2010 to 2011 before the analysis period to avoid post-treatment measurement bias. The "differences p-value" indicates the statistical significance of differences in group means between Conservative LEAs and others in the baseline characteristics controlling for year-fixed effects. The results show that, in the full sample, there are significant differences in most LEA-level characteristics between LEAs controlled by Conservatives and other parties, with many politically "safe" LEA councils. These "safe" LEAs have historically long-standing political control and strong partisan demographic or economic characteristics potentially related to education recommendations. This necessitates using RDD to compare close-election LEAs to compare LEAs with a different timing of switches in political control whose political controls are quasi-random.



| | Full sample | | |
|---|---|---|---|
| | Conservative LEA (N=157) | Others LEA (N=333) | Differences p-value |
| | (1) | (2) | (3) |
| *Panel A. Baseline characteristics* | | | |
| Median age | 40.011 | 38.477 | **0.029** |
| Sex ratio | 0.966 | 0.970 | **0.047** |
| Population density | 1471.500 | 2916.800 | **<0.001** |
| Unemployment rate | 0.320 | 0.373 | **<0.001** |
| White percentage | 0.900 | 0.816 | **<0.001** |
| Life expectancy | 83.863 | 82.380 | **<0.001** |
| Living arrangement as couple | 0.609 | 0.549 | **<0.001** |
| Crime rate: burglary | 944.592 | 1598.784 | **<0.001** |
| Median income | 658.364 | 602.454 | **<0.001** |
| Marriage | 0.482 | 0.419 | **<0.001** |
| Buddhist | 0.611 | 0.579 | **<0.001** |
| Bad general health | 0.046 | 0.062 | **<0.001** |
| With dependent child | 0.286 | 0.298 | **<0.001** |
| No car available | 0.143 | 0.244 | **<0.001** |
| Economically active | 0.838 | 0.793 | **<0.001** |
| Unpaid care | 0.101 | 0.102 | 0.668 |
| Chronic health | 0.157 | 0.180 | **<0.001** |
| Socioeconomic status: high | 0.276 | 0.201 | **<0.001** |

*Note*: P-values describe the significance of differences in group means between Conservative-controlled and Labour-or-others-controlled councils in a linear regression controlling for year fixed-effect.
**Table 3.** Summary descriptive statistics

Figure 1 shows the flexibility of pay recommendations based on LEA geographical location in England in 2017. The map shows that it is possible to either deviate from or recommend flexible pay in different English regions (including London metropolitan borough LEAs). The results help assuage concerns that geographical factors (e.g., regional development, local labor market competitivity) rather than political control affect teacher pay recommendations.



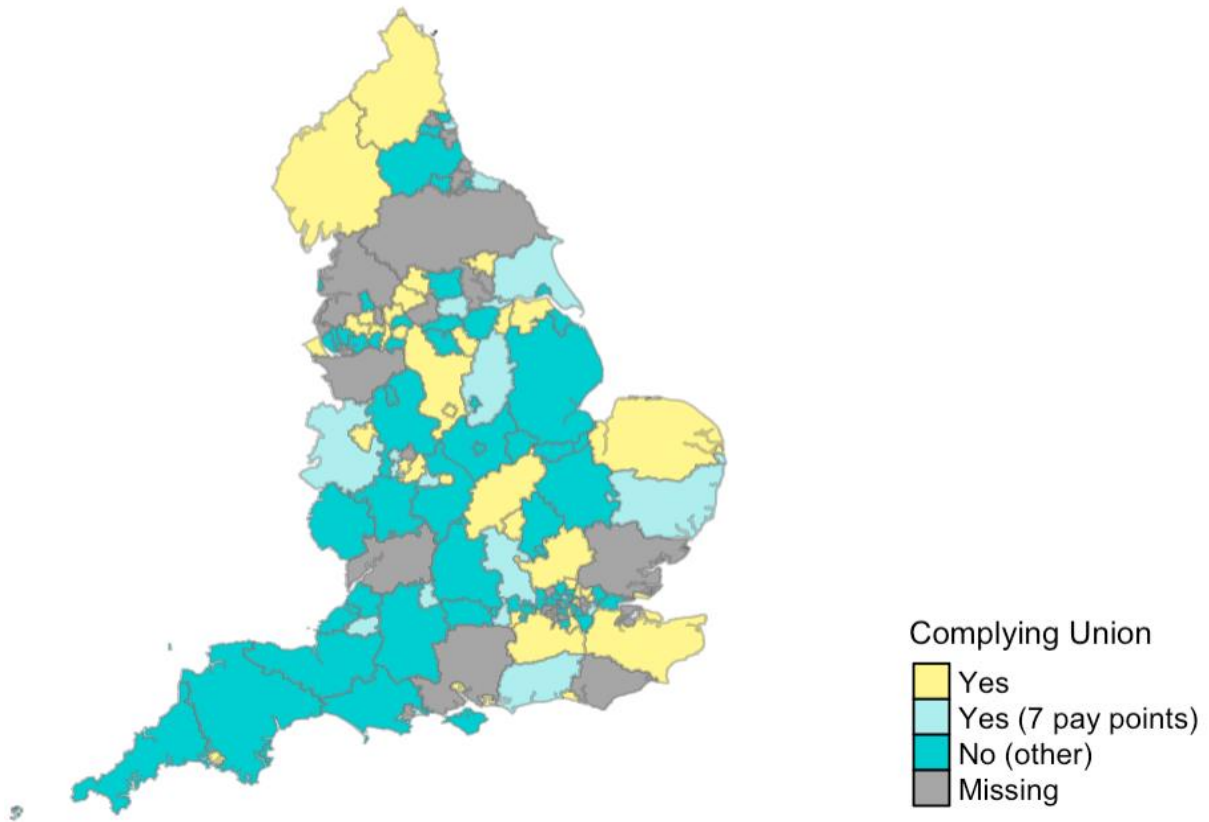

*Figure 1. England LEA compliance with union recommendation in 2017*



# 3. Methodology

## *3.1 Estimation Specification*

To examine the effect of LEA political control, we first compare the naïve differences in group means between elected Conservative-controlled and other-controlled councils in Table 2. However, the party control of LEAs likely correlates with unobservable and observable LEA-level characteristics that could influence teacher pay recommendations, as shown in Table 3. Hence, to identify the causal relationship and avoid endogeneity, we leverage close-election variations in LEA political party control within the RDD.

RDD exploits treatment eligibility criteria based on an observed running variable below or above a cut-off value (Lee & Lemieux, 2014). Under the English electoral system, the LEA governing council's political control changes discontinuously when one party holds 50% or more of the council seats (Alonso & Andrews, 2020). This political setting is different from the US system outlined in Lee (2008), whereby the electoral threshold is not strictly 50% but by the differences in vote share between the strongest and the second-strongest candidate. Specifically, we focus on the marginally Conservative-controlled and other-controlled LEAs (i.e., all non-Conservative parties including Labour, Green, Liberal Democrats, and No-Overall-Control[1]). When the observed continuous running variable, the margin of victory, goes above the 50% majority threshold, it deterministically assigns one LEA-year observation to the Conservative-controlled treatment group and otherwise to the other-controlled control group. It is worth noting that such a "Conservatives-or-others" dichotomy is not the only option. Future studies might adopt more complicated modeling to disentangle the treatment effect by disaggregating "others" into Labour, Green, and Liberal Democrats.

---

[1] There are LEAs where no party has over 50% of the council seats, and the council has no overall control (NOC). These LEAs are assigned as other-controlled LEAs.



Under the assumption that the potential confounders evolve smoothly at the threshold, we compared the quasi-random LEAs just marginally above and below the cut-off to minimize the effect of potential outliers on the causal estimation. RDD has been widely applied to understand the effect of political control on different economic or political outcomes (e.g., Caughey et al., 2017; Lee, 2008; Pettersson-Lidbom, 2012). In a seminal paper by Lee (2008), the author exploits the sharp electoral threshold determining which political party controls the US local congressional office to estimate the incumbency advantage. However, few studies have applied such a research design to investigate the political control effects on substantive social policy-making, especially in non-US contexts.

For the estimation, we use a non-parametric local linear probability model on the binary outcome variable, with recommending schools that aligned with the union being 1 and those that deviated from the union being 0. Logistic regression is not used because the estimators for causal effects with such modeling might suffer from bias and variance trade-offs. The research focus is to understand the local-treatment causal effect within regression discontinuity (RD) bandwidth as opposed to accurately predicting the change probabilities in teacher pay recommendation by LEA party control.

To subset the close-election sample, we use the computationally determined optimal "bandwidth" as described in Calonico et al. (2014). Intuitively, the bandwidth is selected based on data-driven approximations to balance the increasing bias with more observations within the bandwidth and increasing variance with a smaller sample. Specifically, our RD estimation uses an optimal polynomial linear bandwidth $h_{CER}$ determined by minimizing the coverage error rate (CER), which is "preferable in practice" for causal inference (see Calonico et al., 2017, p. 381). In our primary analysis, we do not use the bandwidth algorithm that minimizes the mean square error (MSE) but yields suboptimal confidence intervals by coverage error (Calonico et al., 2019). However, the results of this study are robust to



different bandwidths, as shown in Appendix C. We computationally determine the optimal bandwidth, $h_{CER}$, using the running variable of Conservative vote share margin relative to the 50% seat electoral threshold. Then, we subset the primary full sample of 490 observations to the local close-election sample within the optimal bandwidth and estimate a local linear regression weighted by a triangular kernel. The triangular weighting gives the maximum weights to observations at the majority threshold and declines as the Conservative party council seat margin gets further away. Furthermore, to evaluate the robustness of the results, we estimate the regression with different kernel weighting functions in Appendix C.

Specifically, we estimate the results by employing the following local linear quadratic polynomial approach:

$$Y_{it} = \beta_0 + \beta_1 Conservative_{it} + \beta_2 \widetilde{ConservativeMargin}_{it} + \beta_2 \widetilde{ConservativeMargin}_{it}^2$$
$$+ \beta_3 Conservative_{it} * \widetilde{ConservativeMargin}_{it} + \beta_3 Conservative_{it}$$
$$* \beta_2 \widetilde{ConservativeMargin}_{it}^2 \delta_t + \gamma_i + \epsilon_{it} , (1)$$

where $Y_{it}$, the binary Bernoulli random variable, equals the conditional probability of LEA teacher pay recommendations aligning with the union (i.e., Y=1) in close-election LEA $i$ at year $t$ depending on all covariates. That is, $E\ (Y_{it}) = \frac{1}{n} \sum_{i=1}^{n} Y_{it} = \pi, where\ n$ is sample size within bandwidth. The treatment dummy $Conservative_{it}$ equals 1 for Conservative-controlled LEA and 0 for other-controlled LEA, and the running variable $\widetilde{ConservativeMargin}_{it}$ represents the percentages of Conservatives' vote share $ConShare_{it}$ above or below the electoral victory threshold $c$ (i.e., $\widetilde{ConservativeMargin}_{it} = ConShare_{it} - c, where\ c = 50$).

$\beta_1$ is the parameter of interest capturing the local average treatment effect (LATE) of marginally Conservative-controlled LEAs on the probability of Y=1 by leveraging the quasi-random variations in the close-election LEA party control. LATE refers to the local causal



effect of discontinuous change in the probability of being Conservative-controlled LEA at the majority threshold on the probability of pay recommendations aligning with the union. Given that there is no $X_{it}$ at the threshold that we observe (both treated ($Y_{1it}$) and controlled observations ($Y_{0i}$)), LATE is commonly estimated through the limits of the observations marginally above and below the cut-off, c (i.e., $\tau_{LATE} = E[Y_{1it}|ConShare_{it} = c] - E[Y_{0i}|ConShare_{it} = c] = \lim_{X \to c} E[Y_{it}|ConShare = c] - \lim_{c \to X} E[Y_{it}|Conshare = c]$) (Imbens & Lemieux, 2008).

$\beta_2$ captures the local effects of the Conservative margin of victory in LEA council elections on the flexibility of pay recommendations. $\beta_3$ accounts for any possible difference in this relationship between the LEAs and Conservatives holding overwhelming or marginal LEA political control. The estimation equation also includes only the time-fixed effect $\delta_i$, which allows us to account for the year-specific fixed effect and LEA-fixed effect $\gamma_i$, which accounts for the time-invariant and LEA-specific unobserved factors that might correlate with Conservatives' vote share margin. We follow Ortega (2020) and control for a selection of LEA-level demographic characteristics, $D_i$, including population density, unemployment rate, living arrangements, and bad health rate, to increase the confidence in identifying the causal impact within the RD bandwidth. $D_i$ replaces the LEA-fixed effect in a separate specification, which we describe in Section 3.2. Lastly, $\epsilon_{it}$ accounts for errors uncorrelated across LEAs and time and assumed to have constant variance.

### 3.2 RDD Model Specifications

Four different Equation (1) modeling specifications are estimated within the RD bandwidth. Model (1) examines the pooled association between the treatment, being a Conservative-controlled LEA, and the probability of aligning with the union teacher pay recommendations; no year-fixed effect, LEA-fixed effect, or demographic controls are included (i.e., $\delta_t$, $\gamma_i$, and $D_i$ are constrained to zero). In Model (2), the year-fixed effect is



added. The β coefficients in this model reveal the strength of the association among LEAs of the same year. In Model (3), demographic controls, $D_i$, are added. $D_i$ includes a vector of four covariates, including population density, unemployment rate, living arrangements, and bad health rate, significantly different between Conservative-controlled and other-controlled LEAs within the bandwidth. Reportedly, there is no consistently available demographic measure over the analysis period that covers most LEAs to account for time-varying confounders. Hence, we include both $D_i$ and the year-fixed effect. The results in Model (3) illustrate the associations among LEAs with similar economic and demographic characteristics of the same year. Lastly, in Model (4), we include LEA and year-fixed effects without demographic controls, whose results further strengthen the confidence that the estimation meets the untestable assumption of quasi-randomness within the RD bandwidth. The LEA-fixed effect included in Model (4) picks up any additional unobserved differences among LEAs, beyond the observable characteristics in Model (3). However, it cannot estimate the effect of individual LEA characteristics. In all models, we include population weights to account for high standard errors in the Conservative margin variable of LEAs with smaller populations in all specifications. Thus, the LEAs with smaller populations, smaller council sizes by seats, and hence larger standard errors in Conservative seat margins are given lower weights relative to other large LEAs. We also cluster the standard errors by LEA to address the autocorrelation of errors within LEAs across years.

We test the robustness of the RDD in the following ways. First, we use the balance check to evaluate the differences in demographic and economic characteristics between Conservative-controlled and other-controlled LEAs (Section 4.1). Second, we quantitatively assess whether LEAs sort themselves into or out of treatment status (i.e., Conservative-controlled) using the McCrary test (Section 4.3). Third, we estimate RD regression using different bandwidths and kernel weighting functions (Appendix C). Fourth, we show that the



results remain qualitatively similar when adopting alternative outcome variable definitions or employing similar models that require stronger identification assumptions for causal estimation (Appendix D). Finally, we present the choropleth map that visualizes the geographical distribution of councils by their frequency of appearance in the close-election sample (Appendix E). We discuss the results of these robustness checks in Section 4.3.

*3.3 Alternative Model: Non-Parametric RDD design*

Conventional parametric RDD designs might not fully describe the nature of the data; that is, functional form issues could arise. Herein, we apply nonparametric estimation, which has a flexible functional form. Though it is not the "solution" to the functional form issues, it provides more insights by complementing parametric estimations together. In this study, we use nonparametric kernel regression with a local linear estimator. Let $x_{it}$ denote all covariates; we wish to estimate $Y_{it} = g(x_{it}) + \epsilon_{it}$:

$$Y_{it} = g\big(Conservative_{it}, \widetilde{ConservativeMargin}_{it}, Conservative_{it}$$
$$* \widetilde{ConservativeMargin}_{it}, \gamma_i\big) + \epsilon_{it}. \quad (2)$$

Instead of estimating a regression for all observations in the parametric regressions, local-linear regression estimates a regression for a subset of observations for each point in the data.



# 4. Results

## *4.1 RDD Close-Election Sample*

We mainly use R studio and Stata 16 to perform the analyses. The RDD results are based on the close-election sample of LEAs within the bandwidth, where the Conservatives are marginally winning or losing. The RDD's internal validity relies on the assumption that these close-election LEAs only differ by political control and are otherwise similar in observable characteristics. Any confounding variables evolve smoothly from barely majority Conservative-controlled LEAs to barely other-controlled LEAs. Satisfying this assumption allows the causal inference that any discontinuous variations in the outcome variable can be the unbiased causal effect of political party control.

Although this assumption is empirically untestable on unobservable characteristics, we include the following balance tests to assuage concerns that confounders may bias the estimated LATE maximally. Specifically, we replace the outcome variable in Equation (1) with 17 LEA characteristics to conduct balance checks for statistically significant discontinuities on the observable characteristics between Conservative- or other-controlled LEAs, as seen in Table 4. We compare these statistics for the full sample observations containing 490 LEA-year observations with the "close-election" observations within the computationally determined bandwidth that minimizes CER. In the full sample, Conservative-controlled LEAs are statistically significantly different from other-controlled LEAs in almost every demographic and economic characteristic, as seen in Table 4 (column 3). In the close-election sample, there are four statistically significant differences: Conservative-controlled LEAs have higher population density, higher unemployment rate, more couple co-residence, and worse health. Although we believe these remaining differences between Conservative-controlled and other-controlled LEAs within the bandwidth do not plausibly affect the estimated effect, we include them as covariates in



Model 3 and the LEA-fixed effect in Model 4, which will allow higher precision in RDD estimations. Intuitively, the full-sample LEAs include safe LEA councils where one party has a historical advantage and strong partisan demographic or economic characteristics. The LEAs within the RD bandwidth are the swing councils that could be under control by either party, or the observable characteristics are quasi-random. Hence, the RD estimation results in Models 3 and 4 that compare LEAs with similar characteristics in the close-election sample can even more robustly identify the causal effects of political control alone, the net of other LEA confounders, and support the quasi-randomness of the observations within the RD bandwidth.



| | Full sample | | | | Close election sample | | |
|---|---|---|---|---|---|---|---|
| | Conservative LEA (N=157) (1) | Others LEA (N=333) (2) | Differences in means p-value (3) | | Conservative LEA (N=30) (4) | Others LEA (N=26) (5) | Differences in means p-value (6) |
| Median age | 40.011 | 38.477 | **0.029** | | 41.076 | 42.033 | 0.817 |
| Sex ratio (male/female) | 0.966 | 0.97 | **0.047** | | 0.949 | 0.976 | 0.074 |
| Population density | 1471.5 | 2916.8 | **<0.001** | | 4225.849 | 962.212 | **0.012** |
| Unemployment rate | 0.32 | 0.373 | **<0.001** | | 0.373 | 0.328 | **0.002** |
| White-ethinic percentage | 0.9 | 0.816 | **<0.001** | | 0.82 | 0.816 | 0.207 |
| Life expectancy at birth | 83.863 | 82.38 | **<0.001** | | 84.61 | 83.555 | 0.142 |
| Living arrangement: couple | 0.609 | 0.549 | **<0.001** | | 0.548 | 0.642 | **0.03** |
| Crime rate: homicide | 2.651 | 3.855 | **0.004** | | 3.59 | 2.913 | 0.508 |
| Median income | 658.364 | 602.454 | **<0.001** | | 578.989 | 591.047 | 0.847 |
| Marriage rate | 0.482 | 0.419 | **<0.001** | | 0.459 | 0.488 | 0.164 |
| Christian rate | 0.611 | 0.579 | **<0.001** | | 0.497 | 0.65 | 0.166 |
| General health: bad | 0.046 | 0.062 | **<0.001** | | 0.063 | 0.048 | **0.026** |
| With dependent child rate | 0.286 | 0.298 | **<0.001** | | 0.285 | 0.268 | 0.612 |
| Economically active rate | 0.838 | 0.793 | **<0.001** | | 0.796 | 0.858 | 0.07 |
| Unpaid care rate | 0.101 | 0.102 | 0.668 | | 0.112 | 0.107 | 0.647 |
| Chronic health: limited movement rate | 0.157 | 0.18 | **<0.001** | | 0.187 | 0.173 | 0.592 |
| Socioeconomic status: high | 0.276 | 0.201 | **<0.001** | | 0.209 | 0.173 | 0.808 |
| | | Joint F-statistics | 30.436 | | | Joint F-statistics | 6.927 |

*Note:* All pounds are 2011 British pounds. P-values describe significance of difference in group means between Conservative-controlled and Labour-or-others-controlled councils in a linear regression controlling for year fixed-effect. Close elections include those the margins of victory for the percentage of council seats as the majority is below 6.174% (bandwidth minimizes CER). LEA demographic and economic indicators two years prior to the start of the reform to avoid post-treatment measurements. The joint F-statistics effectively compares the overall goodness of fit of 2 models between one which predicts the Conservative control with none of the LEA characteristics, and one which predicts the Conservative control with all of the LEA characteristics.

**Table 4**. Descriptive baseline characteristics for the full sample and close-election sample



Table 5 shows that the close-election sample also has a similar trend of Conservatives gaining control of more LEAs than the full sample.

**Table 5.** Political Composition statistics for the full sample and close-election sample

| | Full sample | | | | Close election sample | | | |
|---|---|---|---|---|---|---|---|---|
| | **2014** | **2015** | **2016** | **2017** | **2014** | **2015** | **2016** | **2017** |
| Conservative average council seats | 34.33 | 35.034 | 35.537 | 39.013 | 29.579 | 32.305 | 32.821 | 33.792 |
| Labour average council seats | 46 | 46.706 | 46.625 | 44.498 | 50.051 | 49.951 | 49.051 | 49.608 |
| Liberal Democrats average council seats | 9.469 | 8.745 | 8.679 | 8.323 | 10.127 | 6.132 | 6.609 | 5.103 |

*Note:* Close elections include those the margins of victory for the percentage of council seats as the majority is below 6.174% (bandwidth minimizes CER). LEA demographic and economic indicators two years prior to the start of the reform to avoid post-treatment measurements. The elected councilor party affiliation only corresponds to the time when the election took place and, in rare occasions, there are elected politicians defecting their original party but mostly become independent; the descriptive statistics for the primary full sample does not include the excluded LEA-year observations that do not respond to FOI or could not retrieve the pay document; hence, the summary statistics do not fully reflect the England political landscape.

### 4.2 Effects of LEA Party Control on Pay Recommendations

Figure 2 visualizes the RD approach using a scatterplot of the quadratic fits within the bandwidth. The discontinuous jump in the probability of teacher pay recommendations aligning with the union at the electoral threshold already suggests the potential effect of Conservative political control on the contested LEAs. However, without estimating the effects within the close-election bandwidth and controlling for year- and LEA-fixed effects, the discontinuous jump illustrated in Figure 2 might be biased. We also provide the graphical visualization using a nonparametric approach (Figure 3).



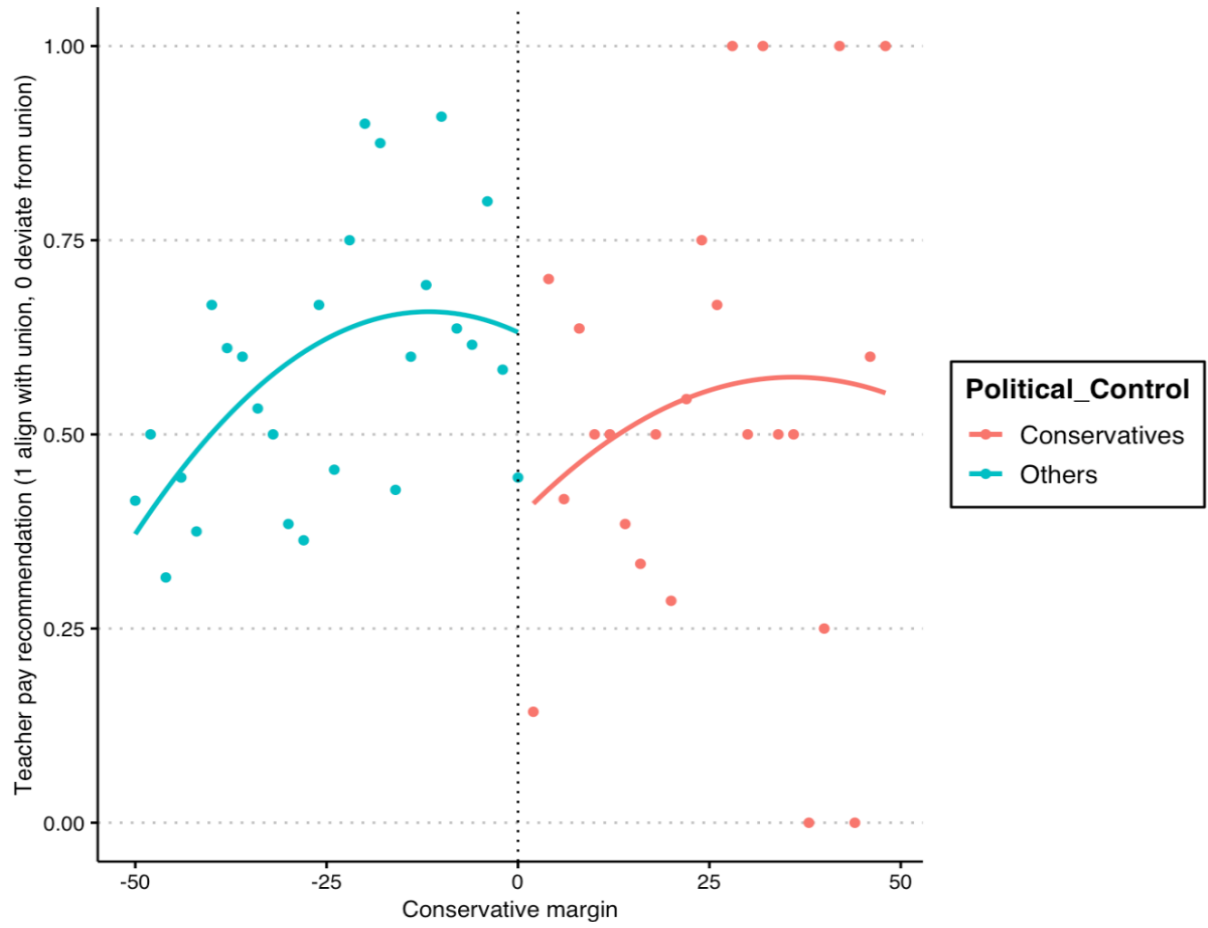

*Figure 2. Scatterplot of RD regression fits*



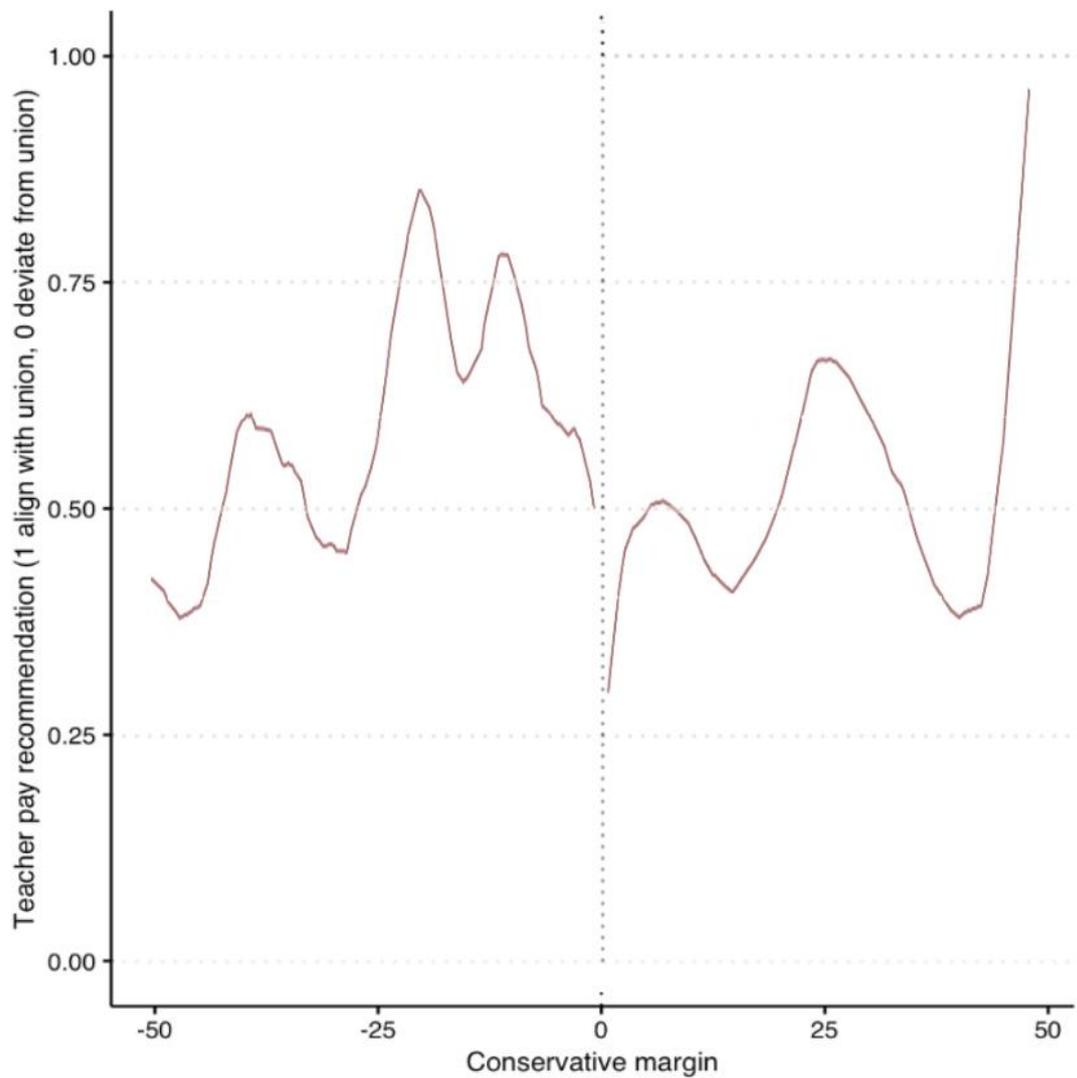

*Figure 3. Mean function of nonparametric regression*

Table 6 shows the baseline probabilities of the teacher pay recommendations in the full and close-election samples. In both samples, the Conservative-controlled LEAs, relative to other-controlled LEAs, are more likely to recommend that schools deviate from union recommendations and adopt performance pay. The contrast in LEA recommendations is even more prominent in the close-election LEAs than in the full sample; 0.533 Conservative-controlled LEAs versus 0.423 other-controlled LEAs recommend that schools deviate from the union. However, without controlling for year- and LEA-fixed effects and adopting triangular kernel weightings for robust estimates, the raw percentage differences shown in Table 6 might be biased.



|  | Full sample | | Close-election sample | |
|---|---|---|---|---|
|  | Conservative LEA (N=157) | Others LEA (N=333) | Conservative LEA (N=30) | Others LEA (N=26) |
|  | (1) | (2) | (4) | (5) |
| *LEA Teacher Pay Recommendations Flexibility (%)* | | | | |
| LEA teacher pay recommendation (follow union recommendation =1, deviate from union recommendation =0) | 0.497 | 0.559 | 0.467 | 0.577 |

*Note:* Close-election samples include those the margins of victory for the percentage of council seats as the majority is above or below 6.174%. The raw percentages do not include triangular kernel weighting, council size weighting, year fixed effect, and the LEA fixed effect.

**Table 6.** Baseline probabilities on the outcome variable

We present the bias-corrected RDD estimates with the robust 95% confidence interval for the local linear probability model in Table 7. The robust estimator presents the unbiased OLS standard errors under heteroscedasticity, whereby the variability of the dependent variable is unequal across the range of predictors. This study primarily uses the combination of bias-corrected and robust estimators, which have been commonly applied in previous RDD studies (e.g., Alonso & Andrews, 2020; Chin & Shi, 2021).



| | | | | |
|---|---|---|---|---|
| **Teacher pay recommendations: align with union =1, deviate from union (i.e., performance pay) =0** | | | | |
| *Models* | *1* | *2* | *3* | *4* |
| Robust Estimates | -0.503** | -0.508** | -0.399*** | -0.512** |
| | [-1.074, 0.069] | [-1.517, -0.178] | [-1.471, -0.331] | [-1.524, -0.190] |
| Year-fixed effect | N | Y | Y | Y |
| LEA-fixed effect | N | N | N | Y |
| LEA Characteristics | N | N | Y | N |
| Clustered SE by LEA | Y | Y | Y | Y |
| Council size weighted | Y | Y | Y | Y |
| N (LEA-year) Left of 0 | 126 | 126 | 126 | 126 |
| N (LEA-year) Right of 0 | 95 | 95 | 95 | 95 |
| BW Left of 0 | 8.794 | 8.549 | 8.615 | 8.581 |
| BW Right of 0 | 8.794 | 8.549 | 8.615 | 8.581 |

*Note:* RD estimates are estimated using local linear regression with triangular kernel function. The bias-corrected robust estimators are developed by Calonico et al., 2017) in the "*rdrobust*" R package. Though "*rdrobust*" does not generate other beta coefficients, we replicate using linear regression without bias correction, triangular kernel, and robust standard errors and report beta coefficients in Appendix G. * p<0.1 ** p<0.05 *** p<0.01. 95% robust confidence intervals are presented in square brackets.

**Table 7.** Impact of Conservative Control on the Probability LEA Teacher Pay Recommendations

The results in Table 7 suggest that marginally Conservative-controlled LEAs are less likely to recommend that schools align with the union and adopt seniority-based teacher pay. Each column displays the $\beta_2$ estimates from Equation (1). In Table 7 (column 1), Model 1's pooled regression suggests that being Conservative-controlled statistically significantly decreases the probability of LEA teacher pay recommendations aligning with the union by 50.3% ($\beta = -0.503$, 95% CI [-1.074, 0.069]).

To check the robustness of the baseline Model 1, we try different model specifications to include sets of covariates such as demographic controls or fixed effects as outlined in Equation 1. Although including covariates is not required for RDD causal identification, it helps increase estimation precision (Imbens & Lemieux, 2008). In Table 7 (column 2), we



include the year-fixed effect to account for by-year variations potentially caused by the following: First, studies suggest that changes in the political landscape of the central government by year during the period are related to local policy-making and political control in England (Clegg, 2021), Romania (Borcan, 2020), and Spain (Solé-Ollé & Sorribas-Navarro, 2008). During the analysis period, the UK central government was led by a coalition of Liberal Democrats and Conservatives in 2014 before Conservatives won with 12 parliamentary seats in the majority from 2015 to 2017. Hence, local policy-making may be affected by such changes in the central government's political control over time. Second, following the 2014 national STRB reform in teacher pay decentralization, local education stakeholders may be in constant and evolving negotiations that might introduce differences in LEA pay recommendations over the years. Willén (2021) studied the short- and long-term effects of the Swedish teacher pay deregulation reform introduced in 1996 and found that the effect of the national reform took three years to fully realize at the local level. Similarly, as the English STRB reform progresses, all local authorities, teachers, and school governing bodies may develop their bargaining skills and adapt to the reform, leading to by-year differences. The Model 2 results suggest that, among LEAs of the same year, being Conservative statistically significantly reduces the probability of teacher pay recommendations being aligned with the union by 33.2% (95% CI [-1.071, 0.052]).

In Table 7 (Model 3), we further include four LEA characteristics significantly different between Conservative-controlled and other-controlled LEAs within the bandwidth. First, the competitiveness of the LEA local labor market may be a crucial factor in educational policy-making. Past research has shown that adopting performance pay in relatively competitive areas leads schools to employ more teachers with lower pay growth and achieve higher student performance (Burgess et al., 2019). To ensure our results are robust to such variations in LEA labor market competitiveness, we include LEA population



density as a proxy for labor competitiveness. Second, we include the unemployment rate, which accounts for local voters' preferences regarding the role of government, since the literature has shown that more deprived populations prefer local governments to provide public services and hence disfavor market-based performance pay (e.g., Brudney et al., 2004; Eardley & Matheson, 2000). Third, living arrangements as a couple and bad health rates capture the LEA demographic structure that affects voters' concerns regarding education policies for the next generation. Recent empirical literature studying the effect of political control on policy-making has extensively applied these covariates in their models (e.g., Alonso & Andrews, 2020; Ortega, 2020). Although including LEA characteristics as covariates in Model 3 leads to a decline in magnitude, the effect of political control remains both substantive and statistically significant at the 95% confidence interval ($\beta$ = -0.530, 95% CI [-0.974, -0.087]).

Table 7 (Model 4) reports a two-way fixed-effect model with the year-fixed effect accounting for longitudinal differences and the LEA-fixed effect as a proxy for all observed and unobserved LEA differences. Intuitively, Model 4 robustly compares the same LEA over time and other LEAs across the same period. It captures any additional unobserved differences among LEAs, beyond the observable characteristics included in Model 3. Compared with the baseline Model 1, the estimated effect size remains similar and corroborates the findings from other models; Conservative-controlled LEAs are statistically significantly 65.0% less likely to align with the union (95% CI [-121.8%, -8.2%]).



**Teacher pay recommendations: align with union =1, deviate from union (i.e., performance pay) =0**

| Models | 1 | 2 | 3 | 4 | 5 | 6 |
|---|---|---|---|---|---|---|
| Robust Estimates | -0.271 ** | -0.327 *** | 0.018 | 0.665 | 0.459 | -0.285*** |
| 95% CI | [-0.479, -0.031] | [-0.534, -0.126] | [-0.529, 0.431] | [-0.152, 1.572] | [-0.289, 1.645] | [-0.482, -0.062] |
| Year-fixed effect | N | Y | Y | N | Y | Y |
| LEA-fixed effect | N | N | N | N | N | N |
| LEA Characteristics | N | N | Y | N | N | Y |
| Council size weighted | Y | Y | Y | Y | Y | Y |
| N | 464 | 474 | 482 | 58 | 81 | 482 |
| Mean BW | 12.170 | 15.069 | 113.119 | 12.207 | 15.065 | 113.119 |
| Kernel | Gaussian | Gaussian | Gaussian | Triangular | Triangular | Triangular |

*Note:* RD estimates are estimated using nonparametric kernel regression. The local-linear kernel estimators are calculated using "npregress kernel" package in Stata. The bootstrap standard errors are calculated with 500 replications (initial "seed" is 3939 for reproducible results). * p<0.1 ** p<0.05 *** p<0.01; 95% confidence intervals are presented in square brackets.

**Table 8.** Impact of Conservative Control on the Probability LEA Teacher Pay Recommendations

The results in Table 8 suggest that marginally Conservative-controlled LEAs are less likely to recommend that schools align with the union and adopt seniority-based teacher pay. Each column displays $\beta_2$ estimates in Equation (1). In Table 8 (column 1), Model 1, the non-parametric regression, suggests that being Conservative controlled statistically significantly decreases the probability of LEA teacher pay recommendations aligning with the union by 27.1% ($\beta$ = -0.271, 95% CI [-0.479, -0.031]). Other kernel and control combinations are also reported in columns 2–6. Model combination (4) in Table 7 is unavailable here because of insufficient degrees of freedom (i.e., valid observations).



*4.3 RDD Robustness Tests*

In addition to the balance test on the 17 demographic characteristics between Conservative-controlled and other-controlled LEAs in Table 4, we test the robustness of the results in the following ways.

RDD assumes continuity in the potential outcomes at the threshold. It is violated if units (i.e., LEAs) can manipulate the values of the running variable, the margin of victory, and sort themselves into or out of treatment (Lee & Lemieux, 2010). In our setting, the assignment manipulation is unlikely within the British local election system since LEAs' councilors should not manipulate voters' choices in the electoral wards. Nonetheless, to confirm the theoretical hypothesis, we use the McCrary test to determine whether the distribution of the treatment is random by looking for bunching in the running variable above or below the cut-off. In Figure 4, we present visual and empirical evidence regarding no discontinuous jumps around the cut-off threshold (McCrary density test significance, p-value=0.226).

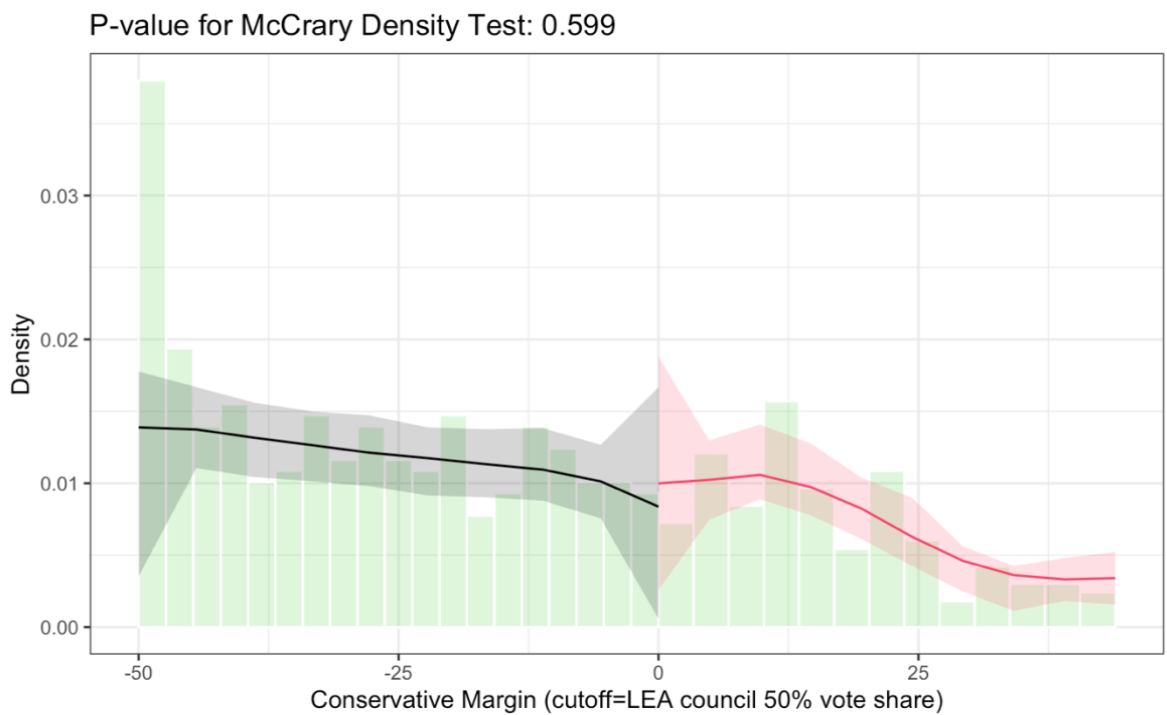



*Figure 4. Distribution of the Conservative margin of victory*

Moreover, Appendix C reports the RD estimates using Model 4 specifications based on different bandwidths from 3 to 50 with step one. The range covers the bandwidths selected by alternative bandwidth selection algorithms: the bandwidth of 8.604, which minimizes MSE (Calonico et al., 2017), and the bandwidth of 3.11 using the algorithm outlined in Imbens and Kalyanaraman (2011). All estimates support our finding that Conservative-controlled LEAs are more likely to deviate from union recommendations and adopt performance pay. We also estimate Model 4 with alternative kernel functions to the triangular kernel weighting function adopted, including the uniform kernel that assigns equal weights and the Epanechnikov kernel that assigns quadratic weightings to the observations within the bandwidth depending on their distance to the cut-off threshold. All estimated effects remain statistically significant and substantively sizable using the three kernel weighting functions.

In Appendix D1, we estimate the model using alternative outcome variable definitions. First, we disaggregate the binary measure of LEA flexibility on teacher recommendations from only 0 and 1 into three-level categorical variables. Instead of collapsing seven-pay-point LEAs into the outcome of 1, we test the association between LEA Conservative political control with three binary dummies of union recommendations: align with the union, deviate from the union (seven-pay-points), and deviate from the union (others). Using linear regression, we re-estimate Equation (1) with each of the three disaggregated binary dummies as outcome variables. The results suggest that Conservative-controlled LEAs are more likely to deviate from union recommendations regardless of the categorization of the seven pay points into aligning or deviating.

Moreover, Equation (1) assumes a non-parametric linear relationship between the probability of recommending schools to align with the union in LEA teacher pay



recommendations and LEA political control by Conservatives. we relax this assumption by presenting an MLR with RDD on the three-level categorical outcome variables in Appendix D2. The results qualitatively support the main finding that, in the close-election sample, the Conservatives are more likely to deviate from union recommendations.

In Appendix E, we present the geographical distribution of councils in the close-election sample and their frequency of appearance in a choropleth map. The plot descriptively suggests that observations evenly located in different parts of England and specific regions should not drive the main findings.



## 5. Conclusion

Decentralized teacher pay matters as an educational policy to promote teacher retention and student performance (Nasim et al., 2022; Willén, 2021). Understanding the political drivers behind this local policy-making serves crucial purposes for academia and future policy designs. After 2014, the STRB reform in England discontinued the central regulation of seniority-based teacher pay while the teachers' union published non-statutory "shadow-STRB" pay schedules. After the reform, the LEAs published recommendations to advise LEA-controlled schools. Given the challenge of balancing the needs of various local stakeholders, LEA councilors rely on their party affiliation to make policy decisions about whether schools should adopt union pay schedules or performance-based flexible teacher pay. This study exploits the STRB reform and quantitatively studies whether Conservative-controlled LEAs are more likely to recommend that schools adopt market-oriented performance pay in the post-reform period.

Existing research evidence for the public policy effects of local political control has been insufficient and inconsistent (e.g., Caughey et al., 2017; Chin & Shi, 2021). Studies focusing on the UK political context have rarely adopted a quantitative quasi-experimental design and focused on central-driven academization (i.e., privatizing state schools) (e.g., Gill & Janmaat, 2019; Hampshire & Bale, 2015). Their findings have qualitatively provided potential evidence of Conservative ideology promoting school autonomy and private-sector influence in local policy-making. Nevertheless, the naïve estimates of the political control effects might be biased by endogeneity issues, whereby political affiliation might reflect voters' ideology, socioeconomic background, and demographic characteristics, which affect local education policies. This study clarifies this research question by providing robust statistical evidence using RDD to identify causal connections. Specifically, we compare the teacher pay recommendations of the marginally winning Conservative-controlled LEAs with



those losing in close elections from 2014 to 2017. Reportedly, this is the first study that explores the immediate and continuing effects of political switches in English policy-making.

Our results from RDD suggest that marginally Conservative-controlled (i.e., right-wing) LEAs are more likely to recommend that schools deviate from teacher union guidance and adopt market-based performance pay. Specifically, being controlled by Conservatives statistically significantly decreases the probability of LEA pay recommendations aligning with the union by 65 percentage points. The findings remain robust when adopting alternative model specifications such as the two-way fixed-effect model or MLR and survive various robustness checks.

Although RDD has numerous advantages in leveraging a quasi-experimental approach to estimate effects, these findings should be interpreted considering our research limitations. First, the RDD estimates LATE with solid internal validity to the local close-election sample but weaker external validity when extrapolating the findings to LEAs with the overwhelming majority. Moreover, England's local politics have a relatively more substantial two-party bias than in other democracies. Hence, to understand the causal mechanism more precisely behind the effects of local political control, future research is required to disentangle local governments' decision-making structures in "safe" LEAs and compare multinational results covering other political institutions with multiple parties.

Second, during the investigation period, the UK's parliamentary control shifted from a Conservative and Liberal Democrat coalition to a Conservative government. Past studies have indicated the potential effect of local and central partisan alignment on policy-making (Clegg, 2021). Further studies could collect data from a period of central political stability during which local government decision-making was less influenced.

Finally, because of a limited number of available observations in the dataset, we cannot further disentangle the effects of political control by different types of teachers pay



recommendations. Future research should disaggregate LEAs by separately considering recommendations that follow the union's six spine points but do not provide pound values and those documents without specific pay recommendations (see Appendix F).

Overall, our results furnish evidence of a local politics and administration dichotomy in England—political control does play a significant role in policymaking. The conclusions from RDD are local to the marginally winning or losing LEAs. However, this study remains relevant in today's political climate with increasing polarization, wherein close elections are becoming increasingly common, and education policies may be more relevant to local politics (Chin & Shi, 2021). This study's results also have wide implications for policy researchers. To date, research on the effects of political control has rarely focused on social policies such as teacher pay decentralization or adopted the quasi-experimental designs in non-US political contexts to offer robust and credible causal identifications. Hence, our study may provide further empirical evidence of politics in social policymaking.


**Acknowledgments**

The authors appreciate Hedvig Horvath, Bilal Nasim, and Alex Bryson for their helpful suggestions and comments, and Aparna Tripathi and Natalie Quinn-Walker for helping to assemble data for the study in collaboration with YL.

**Appendix A. 16 questions used to assess the flexibility of teacher pay recommendations**

Q1. Did the LEA publish a model pay policy for the given year?
   a) Yes, and we have it
   b) Yes, but we do not have it => skip the rest of the questions
   c) No => skip the rest of the questions
   d) Uncertain (we do not know if they published or not) => skip the rest of the questions

Q2. Does the model pay policy contain any explicit recommendations—beyond the min/max range in STPCD—for schools regarding how they should pay their classroom teachers?
   a) Yes
   b) No => skip to Q12

Q3. If the answer to Q2 is yes, does the model pay policy *explicitly* recommend schools follow union recommendations?
   a) Yes => skip to Q12
   b) No

Q4. If the answer to Q3 is no, do recommendations line up with union recommendations?
   a) Yes => skip to Q12
   b) No

Q5. If answer to Q4 is no, are there pay points specified?
   a) Yes
   b) No => skip to Q12

Q6. If answer to Q5 is yes, how many pay points?
   a) Numerical answer expected

Q7. If answer to Q5 is yes, how many of them coincide with union-designated pay points?
   a) Numerical answer expected

Q8. If answer to Q5 is yes, are there any notable pattern in pay points?
   a) Text expected
      ▪ For example, more compressed pay (higher recommendations at the bottom, lower at the top)
      ▪ For example, a bigger jump somewhere?
      ▪ For example, we noticed that some LEAs essentially duplicated the number of main pay scale pay points, where 6 pay points became the union recommendations, and the others the pay points from the previous year (except the lowest one if that fell below the STPCD minimum)
      ▪ Etc.

Q9. If answer to Q5 is yes, please copy and paste them here.
   a) Text expected

Q10. If answer to Q5 is yes, does the model pay policy specify how teachers progress through the pay points?
   a) Yes (if it says e.g., "automatically," "not automatically but upon satisfactory performance," etc.)
   b) No => skip to Q12

Q11. If the answer to Q10 is yes, please copy and paste the relevant word/phrase here.
   a) Text expected ("automatically," "not automatically but upon satisfactory performance," etc.)

Q12. Does the model pay policy suggest pay should be in any way contingent on the teacher's performance/appraisal?



  c) Yes (e.g., by mentioning "pay for performance," "performance pay," "appraisal" somewhere)

  d) No => skip to Q14

Q13. If answer to Q12 is yes, please copy and paste the relevant sentence(s) here.

  ▪ Text expected

Q14. Are there any changes in the structure of pay recommendations between the current year and the previous year?

  a) Yes (mark "yes" also if there was no model pay policy published in the LEA in the previous year)

  b) No

  c) n/a (if there was a model pay policy published in the LEA in the previous year, but we do not have it)

Q15. Month/term the model pay policy is dated to (i.e., not the date when it becomes effective but when it seems to have been agreed/written/published etc.)

  a) MM/YYYY or Term YYYY expected (e.g., 09/2015 or Spring 2015)

Q16. Any notes/comments, anything peculiar you may have noticed in the model pay policy this year?

  a) For example, in some LEAs they are explicitly offered not only for LEA-maintained schools but also for academies.



**Appendix B Renaming of the LEAs in the by-year political composition dataset**

Appendix B1. Renaming process

| Original names | Names for consistency with LEA pay recommendations dataset |
|---|---|
| *Replace "&" with "and"* | |
| Barking & Dagenham | Barking and Dagenham |
| Bath & North East Somerset | Bath and North East Somers |
| Brighton & Hove | Brighton and Hove |
| Cheshire West & Chester | Cheshire West and Chester |
| Hammersmith & Fulham | Hammersmith and Fulham |
| Kensington & Chelsea | Kensington and Chelsea |
| Redcar & Cleveland | Redcar and Cleveland |
| Telford & Wrekin | Telford and Wrekin |
| Windsor & Maidenhead | Windsor and Maidenhead |
| *Clean incorrectly capitalized word "of" and "upon"* | |
| Isle Of Wight | Isle of Wight |
| East Riding Of Yorkshire | East Riding of Yorkshire |
| Kingston Upon Hull | Kingston upon Hull |
| Kingston Upon Thames | Kingston upon Thames |
| Newcastle Upon Tyne | Newcastle upon Tyne |
| Richmond Upon Thames | Richmond upon Thames |
| *Clean the hyphens around 'on'* | |
| Southend-on-Sea | Southend on Sea |
| Stoke On Trent | Stoke-on-Trent |



**Appendix C. RD estimate by different bandwidths and kernel functions**

Using different bandwidths from 5 to 50, the estimates on the RDD model with two-way fixed effect, triangular kernel, weights by council size, and clustered standard errors at the LEA level qualitatively corroborate the findings from the main analytic data that the Conservative-controlled LEAs are more likely deviate from the union recommendations and adopt performance pay. The optimal bandwidth that minimizes CER is colored red. The alternative polynomial bandwidth selection algorithm that minimizes MSE chooses 8.604 as the optimal bandwidth. At that bandwidth of 8.603, the conventional estimate is $\beta = -0.242$ (95% CI [-0.462, -0.023]) and the robust bias-corrected estimate is $\beta = -0.175$ (95% CI [-0.480, 0.131]).

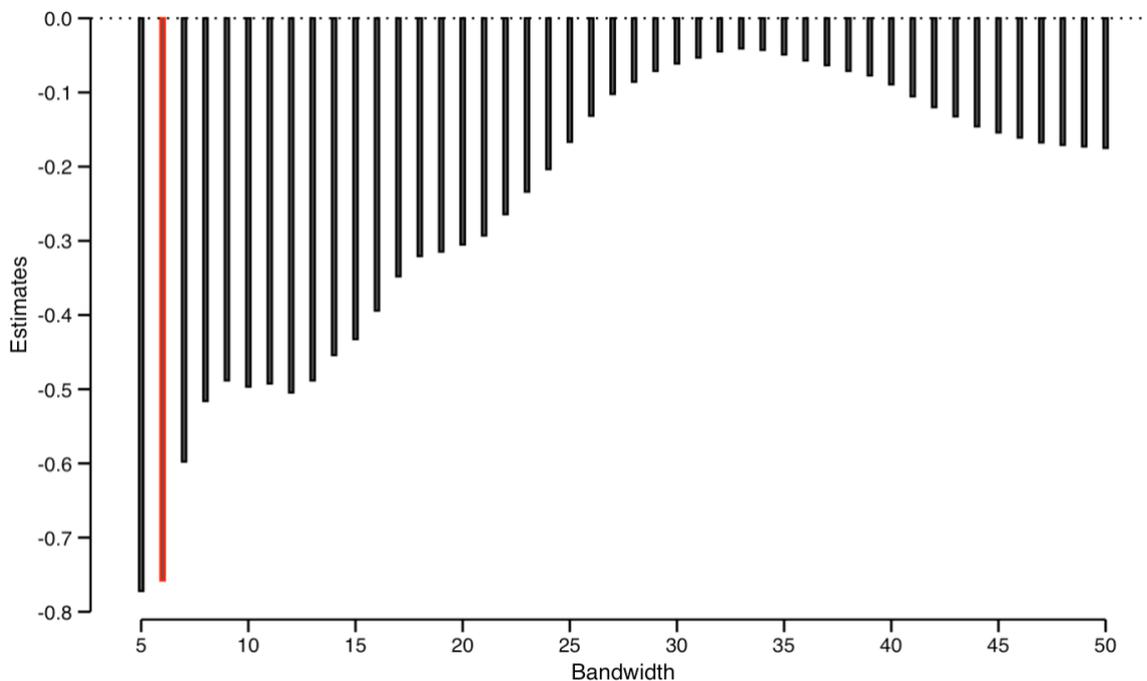

*Figure C1 RD estimates by bandwidths*



*Table C2 RD estimates by kernel*

Kernel functions determine the weighting schemes applied in the non-parametric linear

probability regression and may affect the RD estimates (Cattaneo et al., 2020). In all kernel

functions, observations outside the bandwidth are assigned zero weight. The uniform kernel

treats observations within bandwidth equally. The triangular kernel assigns more weights to

the observations closer to the cut-off and decline linearly as observations with higher absolute

margin of victory. The Epanechnikov kernel assigns similar maximum weights to

observations around threshold but applies quadratic decaying weight.

| Teacher pay recommendations: align with union =1, deviate from union (i.e., performance pay) =0 | | | |
|---|---|---|---|
| | *1* | *2* | *3* |
| Conventional | -0.301 | -0.258 | -0.311 |
| | [-0.751 , 0.148] | [-0.722 , 0.205] | [-0.766 , 0.144] |
| Bias-corrected | -0.546 ** | -0.645*** | -0.650*** |
| | [-0.996 , -0.097] | [-1.109 , -0.181] | [-1.105 , -0.195] |
| Robust | -0.546* | -0.645** | -0.650** |
| | [-1.110 , 0.017] | [-1.250 , -0.041] | [-1.218 , -0.082] |
| Kernel function | Uniform | Epanechnikov | Triangular |
| Year-fixed effect | Y | Y | Y |
| LEA-fixed effect | Y | Y | Y |
| LEA Characteristics | N | N | N |
| Clustered SE by LEA | Y | Y | Y |
| Council size weighted | Y | Y | Y |
| N Left of 0 | 48 | 30 | 30 |
| N Right of 0 | 43 | 33 | 33 |
| BW Left of 0 | 9.508 | 6.671 | 6.727 |
| BW Right of 0 | 9.508 | 6.671 | 6.727 |

*Note:* RD estimates are estimated using local linear regression with different kernel function. The bias-corrected robust estimators are developed by Calonico et al., (2017) in the "*rdrobust*" R package. * $p<0.1$ ** $p<0.05$ *** $p<0.01$. 95% robust confidence intervals are presented in square brackets.

**Table C2.** Impact of Conservative Control on the Probability LEA Teacher Pay

Recommendations by kernel functions



**Appendix D. RD estimates with alternative specifications**

| **Dependent variable classification** | *Deviate =0; 7 pay points and align=1 (the one used in main analysis)* | *Deviate and 7 pay points =0; align=1* | *Deviate and align =0; 7 pay points =1* |
|---|---|---|---|
| Conventional | -0.311 | -0.126 | -0.197 |
| | [-0.766 , 0.144] | [-0.490 , 0.239] | [-0.454 , 0.060] |
| Bias-corrected | -0.650*** | -0.379** | -0.219* |
| | [-1.105 , -0.195] | [-0.743 , -0.015] | [-0.476 , 0.038] |
| Robust | -0.650** | -0.379 | -0.291 |
| | [-1.218 , -0.082] | [-0.921 , 0.162] | [-0.648 , 0.210] |
| Year-fixed effect | Y | Y | Y |
| LEA-fixed effect | Y | Y | Y |
| LEA Characteristics | N | N | N |
| Clustered SE by LEA | Y | Y | Y |
| Council size weighted | Y | Y | Y |
| N Left of 0 | 30 | 41 | 49 |
| N Right of 0 | 33 | 41 | 43 |
| BW Left of 0 | 6.727 | 8.278 | 9.679 |
| BW Right of 0 | 6.727 | 8.278 | 9.679 |

*Notes:* RD estimates are estimated using local linear regression with different kernel function. The bias-corrected robust estimators are developed by Calonico et al., (2017) in the "*rdrobust*" R package. * p<0.1   ** p<0.05   *** p<0.01.  95% robust confidence intervals are presented in square brackets.

**Table D1.** Impact of Conservative Control on the Probability LEA Teacher Pay

Recommendations by outcome variable definition



| Categorical outcome (ref=deviate from union) | Full sample | | Within bandwidth sample | |
|---|---|---|---|---|
| | *7 pay points* | *align with union* | *7 pay points* | *align with union* |
| | *1* | *2* | *3* | *4* |
| intercept | 0 | 3.76*** | 0 | 1.93 |
| | [0.00 − Inf] | [2.03 − 6.96] | [0.00 − Inf] | [0.66 − 5.67] |
| Conservative Seats Margin | 1.01 | 1.03*** | 1 | 0.96 |
| | [0.98 − 1.03] | [1.01 − 1.05] | [0.85 − 1.17] | [0.88 − 1.05] |
| Conservative Won | 0.88 | 0.3*** | 0.93 | 0.75 |
| | [0.30 − 2.61] | [0.13 − 0.70] | [0.11 − 7.88] | [0.20 − 2.89 ] |
| Conservative Won x Conservative Seat Margin | 1.02 | 0.96*** | 1.03 | 1 |
| | [0.98 − 1.03] | [1.01 − 1.05] | [0.85 − 1.17] | [0.88 − 1.05] |
| Year (ref: 2014) | | | | |
| 2015 | 172446140.9 | 0.51** | 37306087.27 | 0.49 |
| | [0.00 − Inf] | [0.29 − 0.90] | [0.00 − Inf] | [0.19 − 1.25] |
| 2016 | 163183674.3 | 0.49** | 23578673.87 | 0.47 |
| | [0.00 − Inf] | [0.28 − 0.86] | [0.00 − Inf] | [0.19 − 1.18] |
| 2017 | 83831752.08 | 0.48*** | 23358673.24 | 0.52 |
| | [0.00 − Inf] | [0.28 − 0.84] | [0.00 − Inf] | [0.21 − 1.32] |



| Observations | 490 | 180 |
| --- | --- | --- |
| R2 McFadden | 0.095 | 0.077 |

*Note:* The bandwidth selection algorithm in Calonico et al., (2017) developed for polynomial linear estimation is suboptimal in nonlinear estimation with the categorical/binary outcome variable. we adopt Xu's (2017) method that provides a robust RD estimator developed for multinomial logistic regression (MLR) under RDD. However, as his estimator does not allow additional covariates, we extracted the computationally determined optimal bandwidth of 17.64497 and estimated the logistic regression using glm and MLR equation using mlogit package in R. Both MLRs are weighted by council size but the standard erors are not clustered due to limited observations and perfect predictions. The reference category is deviated from union recommendations and all presented probabilities are the changes in predicted probabilities when the category is compared to being aligned with union recommendations. * p<0.1   ** p<0.05   *** p<0.01, 95% Confidence Intervals are presented in brackets.

**Table D2.** Impact of Conservative Control on the Probability LEA Teacher Pay Recommendations



**Appendix E. Geographical distribution of observations within the RD bandwidth**

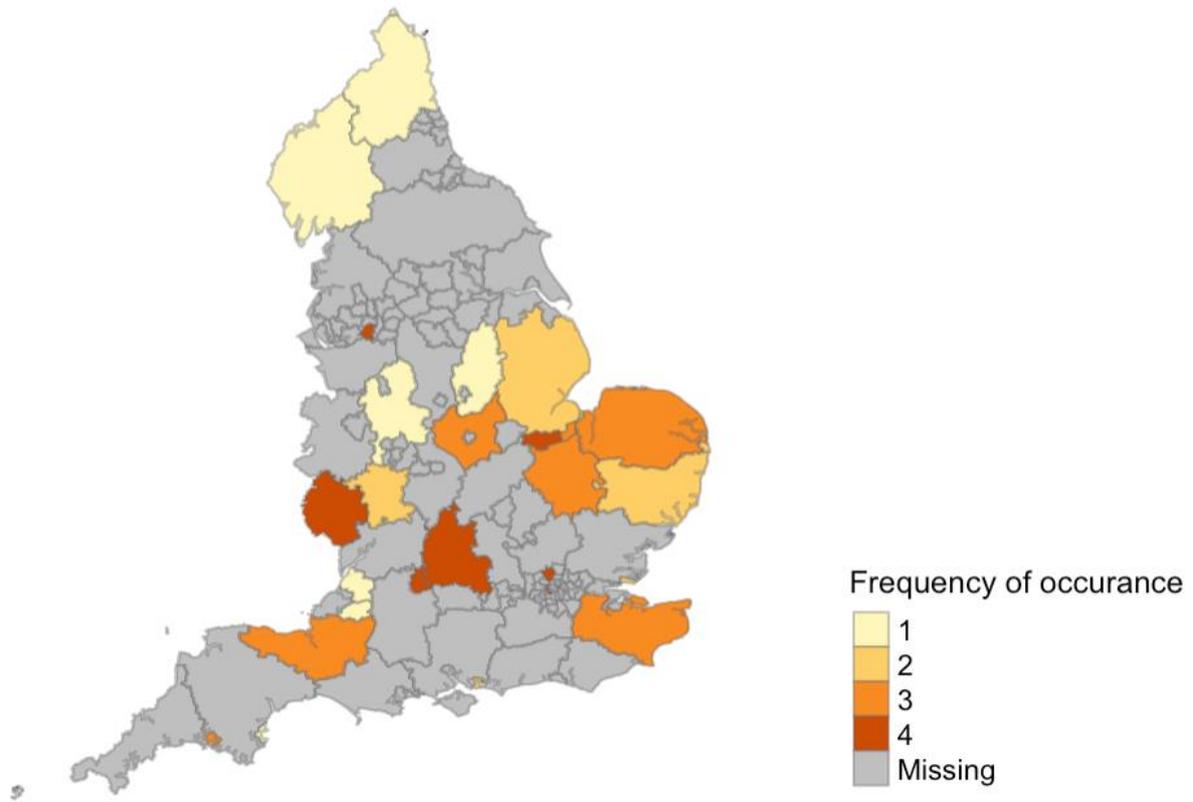

*Note*: The map illustrates the geographical distribution of different LEAs included in the close elections within the RD bandwidth. Missing here means the LEA is not included in the close-election sample.



**Appendix F. Disaggregated measures on teacher pay recommendations**

|  | **2013** | **2014** | **2015** | **2016** | **2017** |
|---|---|---|---|---|---|
| 1. LAs followed union recommendations (including those with at most 1 different point among M2-M5) | 49 | 65 | 38 | 37 | 38 |
| 2. LAs followed 2012 pay points (should only ever be 1 in 2013) | 15 | 0 | 0 | 0 | 0 |
| 3. LAs not coded as 2 and had 6 spine points but no pound values | 6 | 3 | 1 | 1 | 2 |
| 4. LAs not coded as 2 and had 6 spine points but at least 2 different pound values among M2-M5 | 0 | 3 | 1 | 0 | 1 |
| 5. LAs had 7 spine points of the M1-M5-M6a-M6b type (ie. bottom 5 spine points and M6b are in line with union rec, and M6a is the 1% uplifted version of the previous year's M6?) | 0 | 0 | 30 | 30 | 19 |
| 6. LAs providing spine points but more than 6, this category would include those with 7 pay points but not the M1-M5-M6a-M6b regime | 1 | 4 | 6 | 9 | 12 |
| 7. LAs that provided min or max or advised following STRB | 0 | 3 | 4 | 3 | 7 |
| 8.LAs providing pay policy but no recommendations inside | 43 | 39 | 42 | 45 | 47 |
| 9. LAs provided pay policies but we don't have it | 17 | 15 | 11 | 9 | 8 |
| 10. LAs with no answer to FOI request (missing) | 21 | 20 | 19 | 18 | 18 |
| Total observations | 152 | 152 | 152 | 152 | 152 |



**Appendix G. Other RDD coefficients in the local linear regression**

| | | |
|---|---|---|
| intercept | | 0.611 |
| Conservative Seats Margin | | 0.025 |
| Conservative Won | | -0.256 |
| Conservative Won x Conservative Seat Margin | | 0.009 |
| Year (ref: 2014) | | |
| | 2015 | -0.072 |
| | 2016 | -0.069 |
| | 2017 | 0.081 |
| Observations | | 63 |

*Note:* Regression on the observations within the RD bandwidth partially replicates *rdrobust.*

The regression is weighted by council size but does not include triangular weighting function.

The non-robust standard erors are not reported. * p<0.1   ** p<0.05   *** p<0.01